\documentclass[fleqn, twocolumn, amsmath,amssymb,superscriptaddress]{revtex4}
\usepackage{epsfig}
\usepackage{amsmath}
\usepackage{graphicx}
\usepackage{color}
\usepackage{amsfonts,amssymb}
\usepackage{hyperref}
\usepackage{array}
\usepackage{tabu}

\newcommand{\ct}{\cite}
\newcommand{\bi}{\bibitem}
\newcommand{\be}{\begin{equation}}
\newcommand{\ee}{\end{equation}}
\newcommand{\ba}{\begin{eqnarray}}
\newcommand{\ea}{\end{eqnarray}}

\newcommand{\non}{\nonumber}
\newcommand{\bra}[1]{\langle #1|}
\newcommand{\ket}[1]{|#1\rangle}

\begin{document}

\title{Speed and Efficiency Limits of Multilevel Incoherent Heat Engines}
\author{V. Mukherjee}
\affiliation{Department of Chemical Physics, Weizmann Institute of Science, Rehovot 7610001, Israel}

\author{W. Niedenzu}
\affiliation{Department of Chemical Physics, Weizmann Institute of Science, Rehovot 7610001, Israel}

\author{A. G. Kofman}
\affiliation{Department of Chemical Physics, Weizmann Institute of Science, Rehovot 7610001, Israel}
\affiliation{CEMS, RIKEN, Saitama 351-0198, Japan}

\author{G. Kurizki}
\affiliation{Department of Chemical Physics, Weizmann Institute of Science, Rehovot 7610001, Israel}

\begin{abstract}
 We present a comprehensive theory of heat engines (HE) based on a quantum-mechanical ``working fluid'' (WF) with periodically-modulated energy levels. The theory is valid for any  
 periodicity of driving Hamiltonians that commute with themselves at all times and do not induce coherence in the WF. Continuous and stroke cycles arise in opposite limits 
 of this theory, which encompasses hitherto unfamiliar cycle forms, dubbed here hybrid cycles. The theory allows us to  discover the speed, power and efficiency limits attainable by incoherently-operating 
 multilevel HE depending on the 
 cycle form and the dynamical regimes. 
 
\end{abstract}

\maketitle

\section{Introduction}

In recent years, heat engines (HE) comprising quantum mechanical ingredients (the working fluid, baths or work-storing piston/battery) have been a subject of great interest \ct{schwabel06, gemmer10, levy14, curzon75, 
esposito10, campo14, kosloff06, tajima15, wehner15, seifert15, alicki79, kosloff84, geva92, feldmann00, 
rezek06, harbola12, mahler10, linden10, correa13, henrich05, popescu14, quan07, abah14, dorner13, 
dorner12, binder14, malabarba14, seifert16, klimovsky15, klimovsky13, alicki13, klimovsky14, guzik15, uzdin16, chotorlishvili14, chotorlishvili16} 
as part of the broad issue: when can such devices be deemed quantum? And 
if they can, do their performance bounds conform with traditional thermodynamics? Insights into this non-trivial issue first require a good grasp of HE operation
principles whose rapport with quantumness are still
unclear. Such is the dependence of HE performance, i.e., efficiency and power, on the speed (cycle rate) at which they operate and on the scheduling of their coupling to heat baths, which have been 
outstanding issues since the inception
of thermodynamics \ct{schwabel06, gemmer10}. The Carnot cycle, which is a prime example of a ``reciprocating-cycle'' \ct{levy14}, presumes strokes of infinite duration, and 
hence vanishing power. In the Otto cycle, attempts to allow for strokes of finite duration have been primarily confined, for both classical and quantum-mechanical HEs, to
slow operation, as in the Curzon-Ahlborn analysis, which shows that efficiency drops as the speed (cycle rate) increases \ct{curzon75, esposito10, campo14}. Likewise, for a driven 
three-level working fluid (WF) the speed of 
continuous-cycle operation  has been shown to be detrimental, leading to friction, i.e., loss 
of work at the expense of wasted heat production \ct{levy14, curzon75, esposito10, campo14, kosloff06, tajima15, wehner15, seifert15, alicki79, kosloff84, geva92, feldmann00, rezek06}.

Unlike most HE schemes that invoke quantum mechanical working-fluid (WF) systems
\ct{levy14, curzon75, esposito10, campo14, kosloff06, tajima15, wehner15, seifert15, alicki79, kosloff84, geva92, feldmann00, rezek06, harbola12, mahler10,
 linden10, correa13, henrich05, popescu14, quan07, abah14, dorner13, dorner12, binder14, malabarba14, seifert16}, a  minimal HE model based on a periodically modulated qubit
 that is continuously coupled to two spectrally-distinct 
baths actually increases its efficiency with the cycle speed, up to the Carnot bound \ct{klimovsky15, klimovsky13}. The latter bound is reached at the maximal 
speed (modulation rate) that is permissible for HE operation \ct{klimovsky13}. As discussed below, this advantageous performance may be attributed to the frictionless operational regime
of this HE that does not involve any coherence in the WF \ct{klimovsky13, klimovsky15}. 
By contrast, the operation of a HE based on a driven three-level WF \ct{kosloff15}
 crucially depends on the WF coherence (associated with the driving-field action). This difference between the operational regimes of Refs. \ct{klimovsky13} and \ct{kosloff15} implies that WF 
 quantumness is at best optional. 
  Here we do not allow for quantum coherent effects in the WF \ct{scully10, scully13, brumer15}, nor in non-thermal baths \ct{scully03, abah14, hardal15, brumer15, niedenzu15} 
  or in the piston \ct{klimovsky14}.

The considerations outlined above underscore the need for elucidating
the following principle questions: (1) What is the best possible dependence of HE power or efficiency on speed within the Markovian rotating-wave regime? (for non-Markovian or non-rotating wave 
thermodynamic regimes, cf. Refs. \ct{clausen10, shahmoon13, alvarez10, erez08, gordon09}). (2) What is the 
optimal scheduling (cycle form) for attaining the best performance: reciprocating, continuous or possibly some intermediate (hybrid) cycles? Are these cycles equivalent or different in 
terms of performance? (3) Most importantly, is there a fundamental speed limit on HE operation? Insights obtained into these questions will help us resolve the central underlying issue: is 
quantumness essential or advantageous for HE operation?

We address these issues by 
means of a unified theory that applies to any cycle (scheduling) in  multilevel HEs whose driving Hamiltonian commutes with itself at all times and thus does not generate
any quantum coherence in the 
WF: the driving Hamiltonian is diagonal in the energy basis of the WF. Such operation avoids possible friction \ct{levy14, kosloff06, kosloff84, geva92, feldmann00, rezek06}. These HEs are comprised of a
frequency-modulated $N$-level  WF described by, e.g., molecular
angular-momentum giant spin or harmonic-oscillator models, i.e. $2 \leq N < \infty$ (so that the qubit HE model \ct{klimovsky15,klimovsky13} is also included). The WF is subject to
 arbitrary time-dependent periodic coupling to the hot and cold baths, ranging from
continuous coupling in one limit to intermittent coupling and decoupling corresponding to four strokes in an Otto cycle (Sec. \ref{genan}, App. A). Our theory can accommodate diverse reciprocating 
cycles (Sec. \ref{cyclfrm}), 
such as the Otto, Carnot or the two stroke cycles. It allows for a unified treatment
of all possible cycles in the incoherent, Markovian regime (Apps. A-C).
As we show, abrupt (intermittent) on-off coupling to the bath, which is inherent to reciprocating (e.g. Otto) cycles, carries a heavy toll in 
terms of the HE performance, while smoother scheduling is far more advantageous (Sec. \ref{spdlmt}, \ref{spdlmtm}). Finally, we present our conclusions in Sec. \ref{concl}. 
 Insights into the character of the WF steady state and its rapport with thermalization are discussed in App. D.

\section{General analysis} 
\label{genan}

The generic setup (Fig.~\ref{omt}a) is described by the parametrically-modulated Hamiltonian
\ba
H(k,t) = H_{\rm S}(k,t) + H_{\rm I}(k,t) + H_{\rm B}.
\label{hamilsb}
\ea
Here $H_{\rm S}(k,t)$ is the controlled-system (WF) Hamiltonian with modulation period $\tau_{\rm m}$, i.e., 
\ba
H_{\rm S}(k,t + \tau_{\rm m}) = H_{\rm S}(k,t) =  \sum_{n \geq 0} \omega_n(k,t) \ket{n}\bra{n},
\label{hamilsys}
\ea
where
$n$ labels the system levels and the ``smoothness'' parameter $0 \leq k < \infty$ determines the cycle form, ranging from continuous through intermediate 
({\it hybrid}) to reciprocal (stroke) cycle forms (Sec. \ref{cyclfrm}). We have set $\hbar = k_{B} = 1$ for convenience.

As motivated below, there is strong preference to assume $\omega_n (k,t) = n\omega(k,t)$, i.e., to take the levels to be  equidistant and synchronously modulated,
with $k$-independent time average $\overline{\omega_n}(k,t) = n\omega_0$.  Such synchronous modulation of equidistant levels is applicable
to a harmonic oscillator or angular momentum ($N/2$ - spin) WF models \ct{brumer15}, including spin-$1/2$ (two level) WF \ct{klimovsky15, klimovsky13}.
The analysis can be generalized to non-equidistant  level spacings as long as their modulation period $\tau_{\rm m}$
is the same (App. C). More complicated dynamics is beyond the scope of the present work.

The controlled interaction with the 
independent cold (c)
and hot (h) baths is given by
\ba
H_{\rm I}(k,t) = \sum_{j = {\rm c,h}} f_j(k, t) \hat{S}\otimes \hat{B}_j.
\label{hamili}
\ea
Here the operator $\hat{S}$ pertains to a system with an arbitrary number of levels: for angular-momentum models $\hat{S} = \hat{L}_x$ and for a harmonic oscillator $\hat{S} = \hat{X}$ in standard notation
\ct{scully97, cohen77}. 
It is coupled to bath operators $\hat{B}_{\rm c}$ and $\hat{B}_{\rm h}$, satisfying $\left[\hat{B}_{\rm c}, \hat{B}_{\rm h} \right] = 0$, while
$f_{\rm c}(k, t)$ and $f_{\rm h}(k, t)$ are time-dependent system-bath coupling functions parameterized by the smoothness 
parameter $k$. 

It is essential for 
the frictionless dynamics discussed here that the system, and system-bath interaction  Hamiltonians commute with themselves at all times:
 i.e.,  for any $t, t^{\prime}$,
\ba
\left[H_{\rm S}(t), H_{\rm S}(t^{\prime}) \right] = \left[H_{\rm I}(t), H_{\rm I}(t^{\prime})\right] = 0
\label{commi}
\ea

We shall assume an equal or slower periodicity of $H_{\rm I}(t)$ compared to that of $H_{\rm S}(t)$, keeping the two periods commensurate (see below). Under this assumption our 
generalized master equation \ct{breuer02} for the WF density matrix $\rho_{\rm S}(t)$  is (App. A)
\ba
\dot{\rho}(t) = \sum_{j = {\rm c,h}} f_j^2(k, t) \mathcal{L}_j\left[{\rho}(t)\right],
\label{lv1}
\ea
$\mathcal{L}_j$ being the Liouvillian for the $j$th bath, correct to second order in the system-bath coupling. Equation (\ref{lv1}) assumes that the coupling amplitudes
 $f_{j}(k, t)$  are modulated with frequency $\Delta_{\rm I} = 2\pi/\tau_{\rm I}$, which is slow compared to the arbitrary-fast frequency modulation 
 $\Delta_{\rm m} = 2\pi/\tau_{\rm m}$ of $H_{\rm S}(t)$. 

To account for such  modulation, we resort to a Floquet expansion of  the general Liouville 
operator in harmonics of $\Delta_{\rm m}$ \ct{dorner13, klimovsky13, klimovsky15, shahmoon13},  in the rotating-wave approximation (RWA):
\ba
\mathcal{L}_j\left[\rho(t) \right] &=& \frac{1}{2}\sum_{q\geq 0, \omega} G_j(\omega,\pm q) \Big[\hat{S}_{\pm q\omega} \rho(t) \hat{S}^{\dagger}_{\pm q\omega} \non\\
&-& \left(\rho(t) \hat{S}^{\dagger}_{\pm q\omega}\hat{S}_{\pm q\omega} + \hat{S}^{\dagger}_{\pm q\omega}\hat{S}_{\pm q\omega}\rho(t) \right) \Big]\non\\
\label{mastersq}
\ea
Here the raising and lowering operators $\hat{S}^{\dagger}_{\pm q,\omega}$ and $\hat{S}_{\pm q,\omega}$  arise from the expansion of the
system operator $\hat{S}$ (in the interaction picture) in Fourier harmonics $q \geq 0$, as a function of frequency $\omega$: e.g., for a harmonic oscillator, $\hat{S}_{\pm q,\omega}$ is 
related to the annihilation and creation operators (App. A). 
The corresponding Fourier 
component of the $j$th bath 
spectral response (Fourier transform of the bath autocorrelation function),
\ba
G_j(\omega, \pm q) = \int^{\infty}_{-\infty} \langle \hat{B}_j(0) \hat{B}_j(t) \rangle \exp\left[i(\omega \pm  q\Delta_{\rm m})t \right]dt,
\ea 
 becomes  frequency-independent in the Markovian limit \ct{breuer02}. It
fulfills the Kubo-Martin-Schwinger (KMS) detailed-balance condition  
\ba
\frac{G_j(\omega, \pm q)}{G_j(-\omega, \mp q)} =  e^{\beta_j(\omega \pm  q\Delta_{\rm m})},
\ea
$\beta_j$ being the $j$th bath inverse 
temperature. 

The Markovian limit corresponds to $\langle \hat{B}_j(0) \hat{B}_j(t) \rangle \propto \delta(t)$. We note that ``exotic'' 
non-RWA terms (to be considered elsewhere) may give rise to effective squeezing of the system  solely by its extremely fast modulation \ct{shahmoon13}.

In what follows we shall investigate
the QHE performance  in terms of speed limits, efficiency and power, as a function of the modulation rate $\Delta_{m}$ and the cycle form determined by $f_j(k,t)$ and $\omega(k,t)$.

\section{Modelling of cycle forms} 
\label{cyclfrm}

\begin{figure}[t]
\begin{center}
\includegraphics[width= \columnwidth, angle = 0]{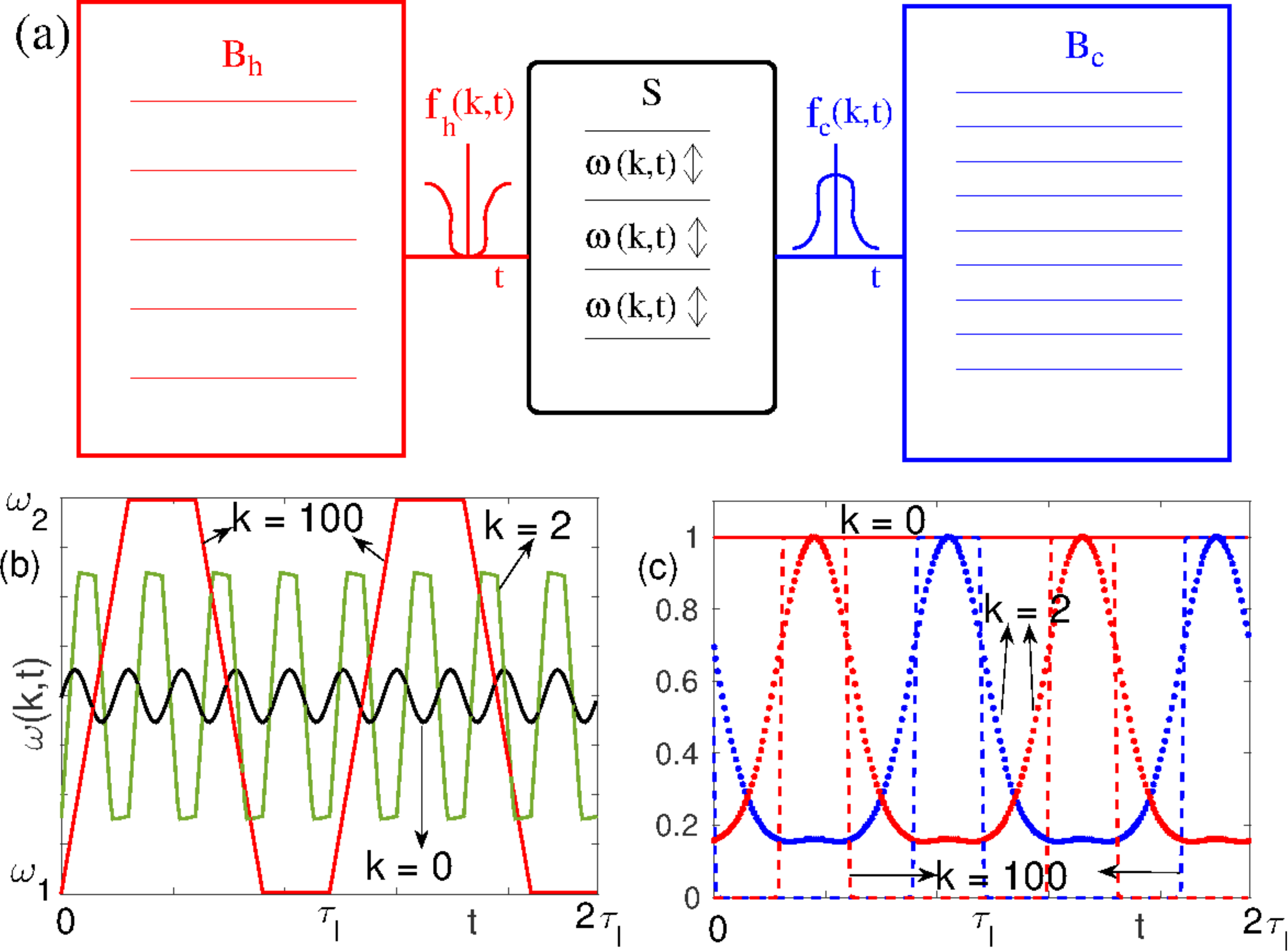}
\end{center}
\caption {(Color Online) (a) Schematic view of the generic HE: multilevel system $S$ with periodic frequency modulation $\omega(k,t)$ and amplitude-modulation of the 
couplings $f_{\rm h}(k, t)$, $f_{\rm c}(k, t)$
to thermal baths
$B_{\rm h}$ (red), $B_{\rm c}$ (blue) with distinct spectra. 
(b) Frequency modulation $\omega(k,t)$: time-variation for different $k$ from continuous to Otto cycles (see text).
(c) $f_{\rm h}(k, t)$ (red) and $f_{\rm c}(k, t)$ (blue) as a
function of time for $k = 0$ (solid), $k = 2$ (dotted) and $k = 100$ (dashed) (see text).}
\label{omt}
\end{figure}

We choose a periodic modulation of $\omega(k,t)$ (Fig. \ref{omt}b) so as to reproduce both the continuous and Otto-cycle limits: 
\ba
\omega(k,t) = \omega_{\rm Cont}(t) \exp[-k] + \omega_0 + \omega_{\rm Otto}(t)\exp[-1/k] ,
~
\label{omteq}
\ea 
where the smoothness parameter $k$ ranges from $0$ to $\infty$.  This parameterization of $\omega(k,t)$ is adopted in order to conform to the $f_j(k,t)$ parameterization (Fig. \ref{omt}c)
discussed below:\\
(a) {\it The continuous-modulation function is chosen to be
\ba
\omega_{\rm Cont}(t) = \lambda \Delta_{\rm m} \sin(\Delta_{\rm m} t),
\label{omcont}
\ea
where $\lambda$ is the modulation depth} \ct{klimovsky13}. 

(b) {\it The function $\omega_{\rm Otto}(t)$ is chosen to be trapezoidal:} this variation characterizes the Otto-cycle limit, where it increases linearly from $\omega_1$ to $\omega_2$ for
$l\tau_{\rm I} < t \leq \ (l + 1/4)\tau_{\rm I}$  (in an isentropic compression stroke), stays at $\omega_2$ until $t = (l + 1/2)\tau_{\rm I}$ (in an isochoric stroke in contact with the hot bath), 
then decreases to $\omega_1$ with the opposite slope 
till $t = (l + 3/4)\tau_{\rm I}$ (in an isentropic expansion stroke) and stays there till $t = (l + 1)\tau_{\rm I}$, where it completes the cycle
(in an isochoric stroke in contact with the cold bath), for a chosen non-negative integer $l$.

In both the continuous ($k = 0$) and the Otto ($k \to \infty$) limits, and hence also for intermediate $k$,  
all levels of the WF must have the same 
modulation frequency $\Delta_{\rm m}$  in order to yield the same sideband spacings $\pm q\Delta_{\rm m}$ (Fig. \ref{phases}a).
Namely, all levels must oscillate synchronously,  which makes the choice of equidistant levels natural (but not compelling).

Our central goal is to find out how do the maximal efficiency and the efficiency at maximal power depend on the operation-cycle form (scheduling), i.e., on the coupling functions $f_j(k, t)$ and on the
speed $\Delta_{\rm m}$.
To this end we parameterize the normalized, periodic $f_j(k, t)$ via the smoothness parameter $0 \leq k < \infty$ ({\bf{\color{red}Fig. \ref{omt}c}}). 
A smooth, hybrid interpolation between the continuous and reciprocal (stroke) cycles corresponds to intermediate values of $k$ for which the coupling to both heat baths is never completely 
switched off or on, and the strokes are not fully separated in time. 
 Accordingly we parameterize the system-bath coupling strengths $f_j(k,t)$ so that they comply with the following requirements: \\
i) {\it periodicity:} $ f_j(k, t + \tau_{\rm I}) = f_j(k, t)$;\\
ii) {\it normalization:} $0\leq |f_j(k, t)| \leq 1$; \\
and\\
iii) {\it variation of $f_j(k, t)$ smoothness with $k \geq 0$.}\\
This parameterization renders a constant 
coupling in the continuous-cycle ($k = 0$)
limit:
\ba
f_{\rm c}(0, t) = f_{\rm h}(0, t) = 1~~\forall~t\non
\ea
 and
stepwise variation in the $k \to \infty$ realistic Otto cycle limit:
\ba
f_{\rm c}(k, t) &=& \theta(t - \tau_{\rm I}/4)\theta(\tau_{\rm I}/2 - t);\non\\
 f_{\rm h}(k, t) &=& \theta(t - 3\tau_{\rm I}/4)\theta(\tau_{\rm I} - t)~~ \text{for} ~ 0 \leq t < \tau_{\rm I}\non,
\ea
 $\theta$ 
being the Heaviside step function.

The period $\tau_I$
of $f_j(k,t)$ is chosen to satisfy
\ba
\tau_{\rm I} = \frac{2\pi}{\Delta_{\rm I}} = \frac{2\pi}{\Delta_{\rm m}}\Phi(k);~~\Phi(k) = \left[\frac{(k + N_1)}{(k + N_2)}\right],
\ea
$\Phi(k)$ being the largest integer closest to the expression in brackets, with $N_1 \gg N_2$. Thus $\Phi(k)\gg 1$ in the continuous-cycle limit $k=0$, and $\Phi(k) = 1$ in the 
Otto-cycle limit $k \to \infty$ where $\Delta_{\rm m} = \Delta_{\rm I}$. We summarize the definitions of the various parameters and functions in Table \ref{tabdef}.
\begin{table}
\caption{Parameters and Functions}
\label{tabdef}
\begin{center}
\begin{tabular}{ | m{6em} | m{6cm}| } 
\hline
~~~ & {\bf ~~~~~~~~~~~~~~~Definitions} \\
\hline
$k$ & Smoothness parameter used to interpolate between a continuous cycle ($k = 0$) and a realistic Otto cycle ($k \to \infty$) (cf.  Figs. \ref{omt}b and \ref{omt}c).  \\ 
\hline
$\omega(k,t)$ & Energy level spacing of the system (cf. Eq. (\ref{omteq}) and  Fig. \ref{omt}b). \\ 
\hline
$f_{\rm {h(c)}}(k,t)$ & Coupling strength of system to hot (h) or cold (c) bath (cf. Fig. \ref{omt}c).  \\ 
\hline
$\Delta_{\rm m} = 2\pi/\tau_{\rm m}$ & Modulation frequency of the system level spacing $\omega(k,t)$. \\
\hline
$\Delta_{\rm I} = 2\pi/\tau_{\rm I}$ & Modulation frequency of the system-bath coupling amplitudes $f_{\rm {h(c)}}(k,t)$. \\
\hline
\end{tabular}
\end{center}
\end{table}
We stress that while the choice of parameterization is arbitrary, 
the behavior it predicts is generic, because any physical cycle form is describable by such parameterization.

\section {Heat flow, work and efficiency}
\label{spdlmt}

With the above ``smoothness'' $k$-parametrization at hand, the heat currents $J_{\rm c}$ and $J_{\rm h}$, flowing out of the cold and hot baths, respectively, are obtained 
from eqs. (\ref{lv1})
consistently with the Second Law \ct{klimovsky13, klimovsky15} in the form 
\ba
&&J_{\rm h}(t)\non\\
&&= f_{\rm h}^2(k, t) \sum_{q \geq 0}(\omega_{0} + q\Delta_{\rm m}) P_q(k) G_{\rm h} (\omega_0 + q\Delta_{\rm m}) \mathcal{F}_{\rm h}(q, k,t) \non\\
&&J_{\rm c}(t)\non\\
&&= f_{\rm c}^2(k, t) \sum_{q \geq 0}(\omega_{0} - q\Delta_{\rm m}) P_q(k) G_{\rm c} (\omega_0 - q\Delta_{\rm m}) \mathcal{F}_{\rm c}(q, k,t)\non\\
~
\label{Jhc}
\ea
The harmonic (sideband) weights are denoted by $P_q(k)$, $G_{\rm h(c)}(\omega_0 \pm q\Delta_{\rm m}) \equiv G_{\rm h(c)}(\omega, \pm q) $, while $\mathcal{F}_{\rm h(c)}(k,t)$ 
express the h(c) contributions to the detailed balance between heat emission and absorption (see details in App. C). 
\begin{figure}[t]
\begin{center}
\includegraphics[width= 0.77\columnwidth, angle = 0]{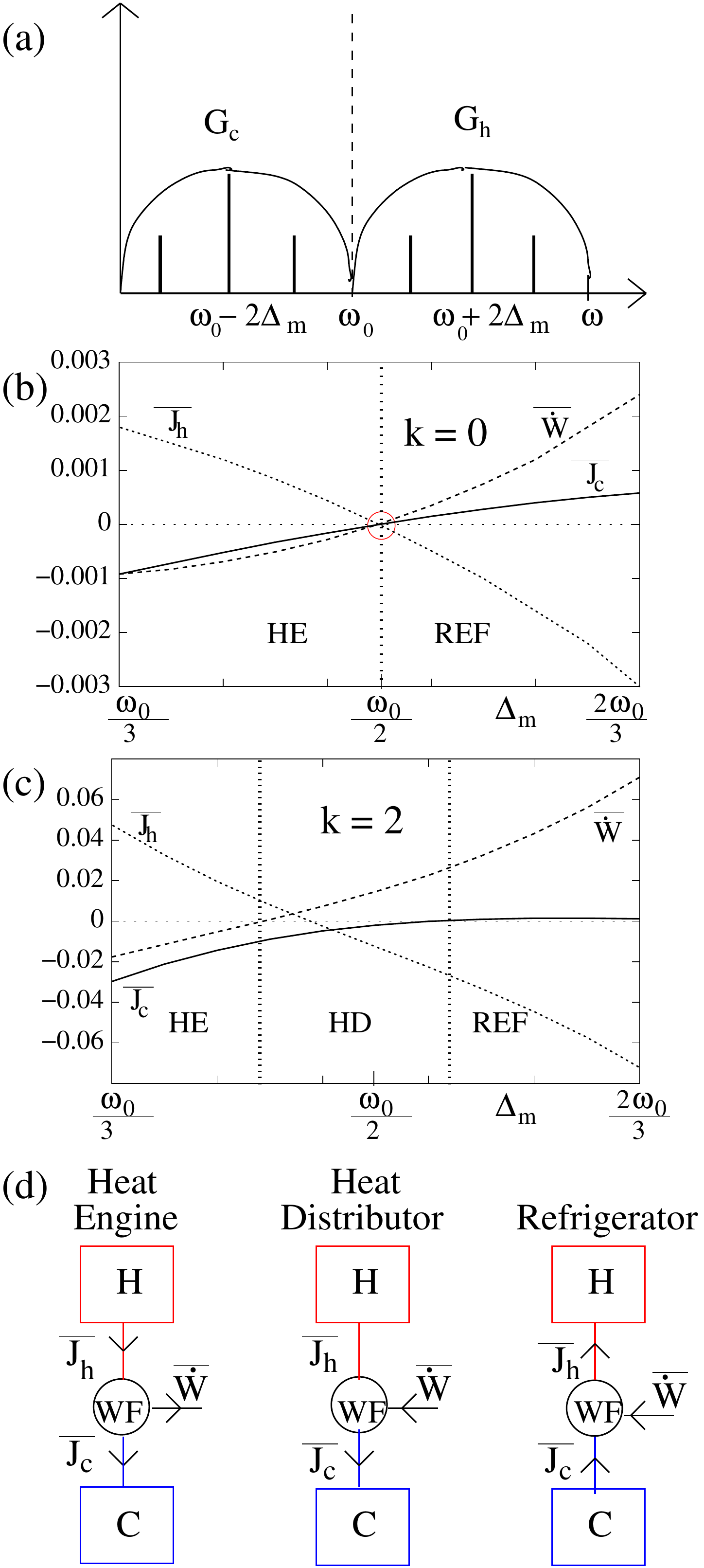}
\end{center}
\caption {(Color Online) (a) Spectral representation of the hot and cold baths and frequency modulation sidebands in the Markovian limit. 
(b) Heat currents and power for the continuous 
cycle ($k = 0$) yield two possible regimes:
HE ($\overline{\dot{W}} < 0, \overline{J}_{\rm c} < 0, \overline{J}_{\rm h} > 0$) for $\Delta_{\rm m} < \Delta_{\rm cr}$  and
refrigerator (REF) ($\overline{\dot{W}} > 0, \overline{J}_{\rm c} > 0,  \overline{J}_{\rm h} < 0$) for $\Delta_{\rm m} > \Delta_{\rm cr}$ (cf. Eq. (\ref{delcr})). For our choice of parameters, 
$\Delta_{\rm cr} = \omega_0/2$.
(c) Same for a hybrid cycle ($k = 2$). Now we have three regimes: HE, heat distributor 
(HD) ($\overline{\dot{W}} > 0,  \overline{J}_{\rm c} < 0, \overline{J}_{\rm h} \in \mathbb{R}$) and 
refrigerator. 
Here $T_{\rm c} = 10$, $T_{\rm h} = 30, \omega_0 = 3, \lambda = 0.1, G_{\rm c}(0< \omega_0 - q\Delta_{\rm m} < \omega_0) = G_{\rm h}(\omega_0 + q\Delta_{\rm m} > \omega_0) = 1$.
(d) Schematic operational regimes of HE, heat distributor 
 and refrigerator. }
\label{phases}
\end{figure}
\begin{figure}[t]
\begin{center}
\includegraphics[width= 1.05\columnwidth, angle = 0]{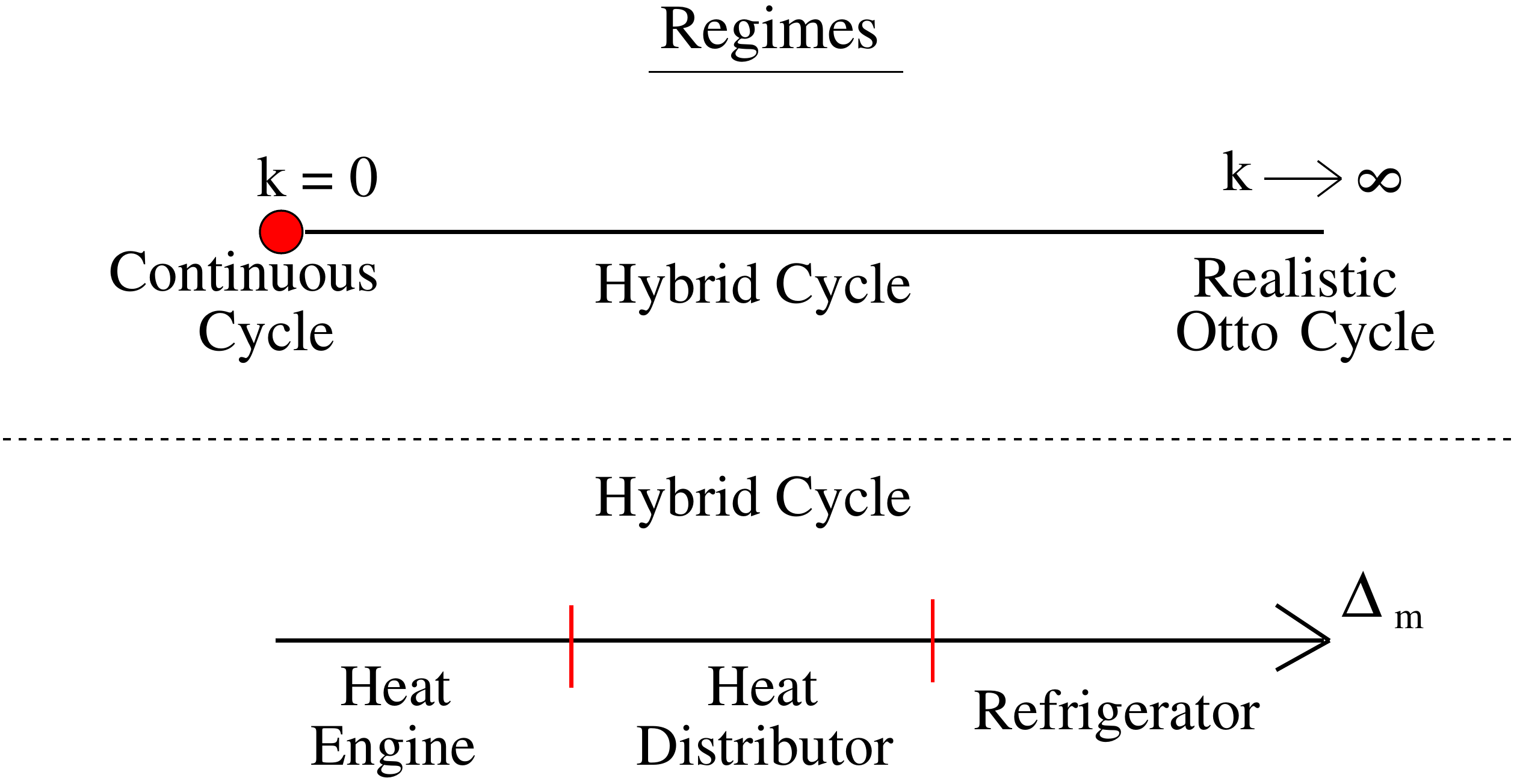}
\end{center}
\caption {(Color Online) A continuous cycle (red dot) is obtained for $k = 0$, whereas $k \to \infty$ describes a realistic Otto cycle. 
For a hybrid cycle, the heat machine operates in the heat engine, heat distributor or refrigerator regimes depending on the modulation rate $\Delta_{\rm m}$. 
The heat distributor regime is absent for the continuous cycle $k = 0$.}
\label{regms}
\end{figure}

A crucial condition of our treatment of diverse (hybrid) cycles, and their 
continuous and stroke cycle limits, is the spectral separation 
of the hot and cold baths, such that the sidebands $\pm\left(\omega + q\Delta_{\rm m}\right)$ only couple to the hot bath and the
$\pm\left(\omega - q\Delta_{\rm m}\right)$ sidebands  only couple to the cold bath, as detailed below (Fig. \ref{phases}a). This spectral separation,  which is compatible with the Markovian limit,
is required to allow selective control of the heat currents, 
which is the essence of HE operation.
In order to allow for HE operation, we require that positive
($\omega_0 + q\Delta_{\rm m}$) or negative
($\omega_0 - q\Delta_{\rm m}$) sidebands be non-vanishing in $\mathcal{F}_{\rm h}$ and $ \mathcal{F}_{\rm c}$ respectively, thereby controlling the heat flow sign (direction).
This requirement amounts to
\ba
&&\mathcal{F}_{\rm h} > 0,~ \mathcal{F}_{\rm c} < 0,\\ 
&&G_{\rm h}\left(\omega_0 - q\Delta_{\rm m} \leq \omega_0 \right) \approx 0;~~G_{\rm c}\left(\omega_0 + q\Delta_{\rm m} \geq \omega_0\right) \approx 0.\non
\ea
This equation implies a separation of the spectral couplings to the two baths for all contributing harmonics (Fig. \ref{phases}a).

According to the First Law of thermodynamics for a parametrically driven $H(t)$, the power $\dot{W}$ is given by \ct{klimovsky15, alicki79} 
\ba
\dot{W}(t) = -(J_{\rm h}(t) + J_{\rm c}(t)).
\ea
The possible operational regimes of the heat machine are: HE ($\overline{\dot{W}} < 0, \overline{J}_{\rm c} < 0, \overline{J}_{\rm h} > 0$), 
heat distributor
($\overline{\dot{W}} > 0, \overline{J}_{\rm c} < 0, \overline{J}_{\rm h} \in \mathbb{R}$) and refrigerator ($\overline{\dot{W}} > 0, \overline{J}_{\rm c} > 0, \overline{J}_{\rm h} < 0$). 
The occurrence of each regime is
 determined by the signs of  cycle-averaged $\overline{J}_{\rm h}$, $\overline{J}_{\rm c}$ amd $\overline{\dot{W}}$, as shown in Figs. (\ref{phases}) 
and (\ref{regms}). 

We may use Eq. (\ref{Jhc}) to calculate the steady-state efficiency $\eta$  and cycle-averaged power output 
\ba
\eta = -\frac{\oint_{\tau_{\rm I}} \dot{W}(t) dt}{\oint_{\tau_{\rm I}} J_{\rm h}(t) dt};~~~ \overline{\dot{W}} = \frac{1}{\tau_{\rm I}} \oint_{\tau_{\rm I}} \dot{W}(t) dt
\label{effpw}
\ea
as a function of the  modulation speed $\Delta_{\rm m}$, the cycle duration $\tau_{\rm I}$, and of the smoothness parameter $k$, searching for the maxima
of the functions in Eq. (\ref{effpw}) (see Figs. \ref{eff_Del} and \ref{effk}). The efficiency $\eta$ is here defined for the HE regime, wherein $\overline{\dot{W}} < 0$.

As mentioned before, the performance bounds embodied 
by $\eta$ and $\overline{\dot{W}}$ (cf. Figs. \ref{eff_Del} and \ref{effk}) do not include any quantum coherence-related features: although the 
expressions reflect the quantized WF level structure, they have
classical counterparts (see Discussion, Sec \ref{concl}).

\section{Speed-limit for hybrid cycles}
\label{spdlmtm}

\subsection{General Speed Limits} 

\begin{figure}[t]
\begin{center}
\includegraphics[width= 0.95\columnwidth, angle = 0]{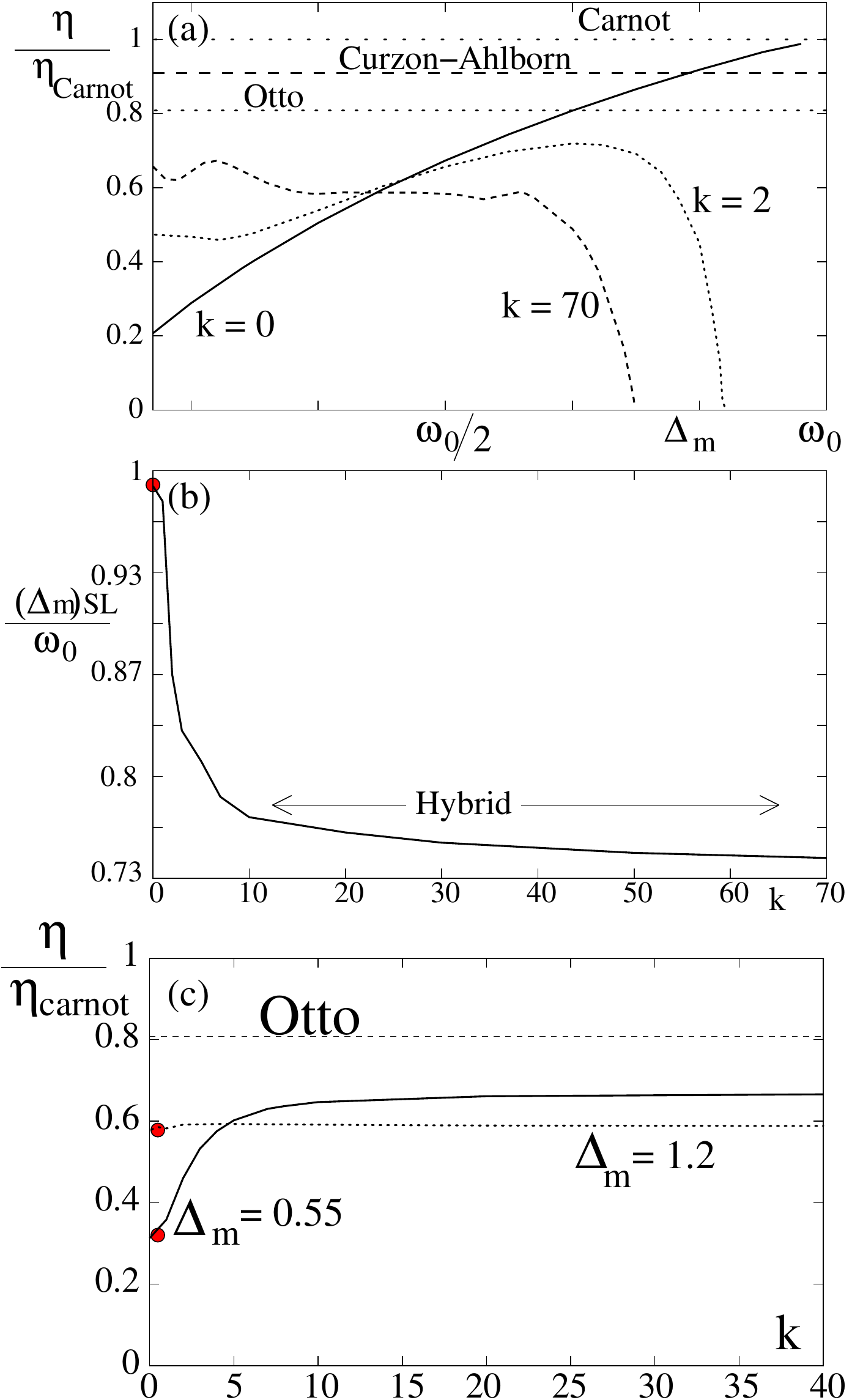}
\end{center}
\caption {(Color Online) (a) Efficiency $\eta$ scaled by $\eta_{\rm Carnot}$ as a function of modulation rate $\Delta_{\rm m}$ for the different $k$ values. 
$\eta(k = 0)$ may exceed the 
Curzon-Ahlborn bound, as well as the Otto-cycle efficiency $\eta_{\rm Otto}$,
whereas cycles with larger $k$ fail to do so.  Our choice of parameters yields $\Delta_{\rm cr} \approx 0.98\omega_0$, thus signifying that
for the continuous $k = 0$ cycle, the WF operates in the refrigerator regime only for very large $\Delta_{\rm m}$, not visible in the range plotted above. 
(b) Speed limit: $\left(\Delta_{\rm m}\right)_{\rm SL}$ as a function of $k$, same parameters as in (a). Continuous-cycle limit: red dot. 
(c): Efficiency $\eta$ scaled by $\eta_{\rm Carnot}$ as a function of $k$ (same parameters as above): $\eta$ tends to 
 $\eta_{\rm Otto}$ as $\Delta_{\rm m}$ decreases. Red dots show 
the efficiencies for the continuous-cycle. (Here 
$\omega_1 = 1, \omega_2 = 5, \omega_0 = 3, T_{\rm c} = 1, T_{\rm h} = 100$, $G_{\rm c}(0< \omega_0 - q\Delta_{\rm m} < \omega_0) = G_{\rm h}(\omega_0 + q\Delta_{\rm m} > \omega_0) = 1$).}
\label{eff_Del}
\end{figure}

In the fast-modulation (large $\Delta_{\rm m}$) range of a  $k = 0$ (continuous) cycle, where $P_{q>1} \ll P_{q=1}$, the onset of the refrigerator regime ($\overline{\dot{W}} > 0$, $\overline{J_c} > 0$)
occurs when
\ba
\Delta_{\rm m} > \Delta_{\rm cr} = \omega_0 \frac{T_{\rm h} - T_{\rm c}}{T_{\rm h} + T_{\rm c}}.
\label{delcr}
\ea
This is the same condition as in the minimal (two-level) WF model \ct{klimovsky13}. In such a cycle, one can control the speed limit 
which corresponds to the cross-over from the HE to the refrigerator regime. We may then vary $\Delta_{\rm cr}$ by tuning $T_{\rm h}$ and 
$T_{\rm c}$ only, as shown in Fig. \ref{phases}b ($\Delta_{\rm cr} = \omega_0/2$) and Fig. \ref{eff_Del}a ($\Delta_{\rm cr} \approx 0.98 \omega_0$) 
\begin{figure}[t]
\begin{center}
\includegraphics[width= 0.95\columnwidth, angle = 0]{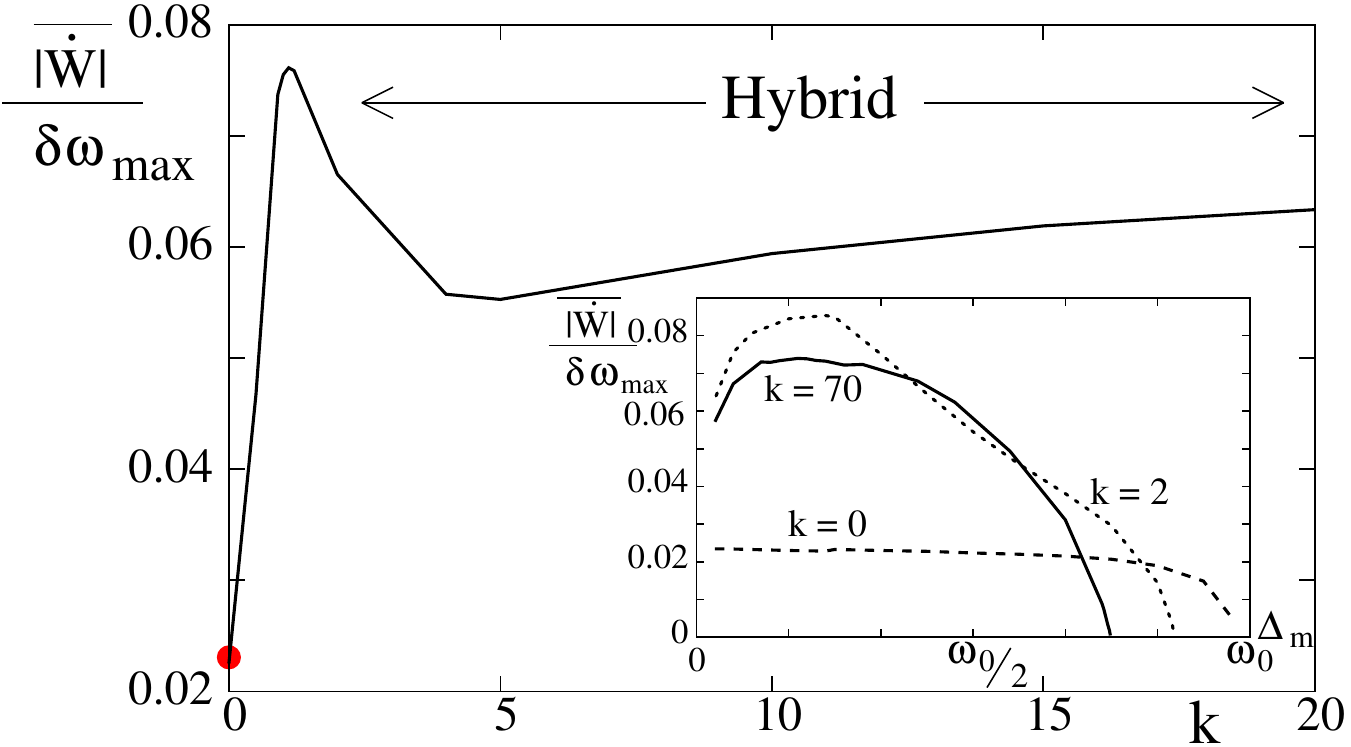}
\end{center}
\caption {(Color Online)  Absolute generated average power $|\overline{\dot{W}}|$ scaled by $\delta \omega_{\rm max}$ as a function of $k$.
$\delta \omega_{\rm max} (k)$ is the
maximum  modulation amplitude of $\omega(k,t)$, which is related to the input power. 
A hybrid HE cycle ($k \approx 1.1$) yields a higher power
than either a continuous or a realistic Otto cycle with $k \to \infty$. 
 The continuous-cycle limit is shown by the red dot. (Here $\Delta_{\rm m} = 1.2$)
Inset: Same as a function of $\Delta_{\rm m}$. In the limit of
$\Delta_{\rm m} \to 0$ (and $\tau_{\rm I} \to \infty$), 
the average power vanishes for large $k$, as indicated in the plot. 
(Same parameters as in Fig. \ref{eff_Del}).}
\label{effk}
\end{figure}

In contrast, an additional intriguing regime arises only for finite $k$ when the cold bath interacts with effective \emph{negative frequencies} of system, 
resulting in a ``heat distributor''(HD) regime, 
in which 
\ba
\omega_0 - q\Delta_{\rm m} &<& 0 \non\\
\overline{\dot{W}} > 0,~~ \overline{J_{\rm c}} &<& 0,
\ea
 indicating that (positive) work is done \emph{on} the WF, which in turn transfers heat to the cold bath (see Fig. \ref{phases}). 
The modes with 
large $q$ may then 
contribute either to the refrigeration of the cold bath, or to the HD that does not produce useful work,  whereas modes
corresponding to smaller $q$'s still contribute to the HE regime. For large $k$, the refrigerating and heat distributing modes
 decrease the efficiency and power to $\eta \to 0$ and $\overline{\dot{W}} \to 0$ (see Figs. \ref{eff_Del} and \ref{effk}) as
$\Delta_{\rm m}$ approaches  the speed limit $\left(\Delta_{\rm m}\right)_{\rm SL}$, which is bounded by
\ba
\left(\Delta_{\rm m}\right)_{\rm SL} \leq \Delta_{\rm cr}.
\label{omcr}
\ea
The equality $\left(\Delta_{\rm m}\right)_{\rm SL} = \Delta_{\rm cr}$ only holds for $k = 0$, i.e., for a continuous cycle. However, $\overline{\dot{W}} = \overline{J_h} = 0$ 
at $\Delta_{\rm m} = \Delta_{\rm cr}$ for a 
 continuous cycle (see Fig. \ref{phases}b), so that $\eta$ becomes ill-defined.

\subsection{Speed Limits from Continuous to Otto Cycles}
 Traditionally, in an Otto cycle 
the WF is translationally displaced between the strokes, so that it may intermittently couple to the hot bath at frequency $\omega_2$ and to the cold bath at frequency $\omega_1$. Here, instead,
we aim to reproduce any cycle, including a (slightly-smoothed) approximation to the Otto cycle, without physically moving the WF, but rather by spectral separation of the couplings to the two baths, 
under an appropriate choice of
the modulation harmonics $q\Delta_{\rm m}$. 

The abrupt on-off switching of the strokes in a traditional Otto cycle (which we dub TOC) 
is not only idealised, but also entails friction,  which
is difficult to overcome \ct{campo14}. By contrast, a frictionless realistic Otto-cycle (which we dub ROC)  is reproduced by allowing a large number of harmonics to become 
significant as $\Delta_m \to 0$, such that the hot bath effectively couples to the WF only 
at 
\ba
\omega_2 = \omega_0 + q_{\rm Otto}\Delta_{\rm m}
\label{om2ot}
\ea
and the 
cold bath at 
\ba
\omega_1 = \omega_0 - q_{\rm Otto}\Delta_{\rm m}.
\label{om1ot}
\ea
As we show, ROC fundamentally differs from TOC:  Eqs. (\ref{om2ot}), (\ref{om1ot}) impose a speed limit of HE
operation on ROC, which is in general absent in a ``perfect" TOC; the latter has no speed limit.

The above discussion allows us to answer question
(3) in the Introduction:
There is indeed a speed limit  for any realistic cycle, including ROC, in the sense that a modulation rate $\Delta_{\rm m}$ above $\left(\Delta_{\rm m}\right)_{\rm SL}$ results
in the system acting as a heat distributor, or a
refrigerator of the cold bath, and thus consuming, rather than generating, 
power: $\overline{\dot{W}} > 0$. 
Yet the speed limit $\left(\Delta_{\rm m}\right)_{\rm SL}$ depends on scheduling, i.e., it decreases with 
increasing $k$. The highest speed limit is compatible with a continuous cycle. By contrast, a  frictionless, finite-duration ROC demands an increasingly slower modulation in order to produce 
power ($\overline{\dot{W}} < 0$) (Fig. \ref{eff_Del}b). 

The efficiency matches the value 
\ba
\eta_{\rm Cont} = \frac{2\Delta_{\rm m}}{(\omega_0 + \Delta_{\rm m})}
\ea
in the continuous limit $k \to 0$, where only the first harmonic (sideband) is significant \ct{klimovsky13}. This expression is bounded by the Carnot efficiency:
 The Curzon-Ahlborn limit may be surpassed and $\eta_{\rm Carnot}$ is attained in the continuous limit. For large $k$,
$\eta(k)$ saturates to the lower Otto-cycle efficiency 
\ba
\eta_{\rm Otto}(k \to \infty) = 1 - \frac{\omega_1}{\omega_2}
\ea
when the frictionless ROC satisfies 
\ba 
q_{\rm max} &\to& \infty,~~\Delta_{\rm m} \to 0,\non\\
q_{\rm max} \Delta_{\rm m} &\lesssim& \Delta_{\rm cr}
\ea
as per Eq. (\ref{omcr}), keeping the corresponding power $\overline{\dot{W}} < 0$.

Remarkably, our results suggest that the maximal power is attained near the 
$k$ value where $\omega(k,t)$ (Eq. (\ref{omteq})) changes from $\omega(k,t) \approx \omega_{\rm Cont}(t) + \omega_0$ to $\omega(k,t) \approx \omega_{\rm Otto}(t) + \omega_0$. 
We thereby reveal the possibility of engineering a hybrid-cycle HE which outperforms both the continuous and the Otto limits (Fig. \ref{effk}) because of its optimal 
cost of coupling to and decoupling from 
the bath.

We are now in a position to answer the principle
questions (1) and (2) in the Introduction: the optimal (best) cycle form (scheduling of an incoherent HE) is a hybrid cycle with $0 < k <\infty$ for 
which the scaled power $|\overline{\dot{W}}| /\delta \omega_{\rm max}(k)$ peaks ($k \approx 1.1$ for the chosen parameterization). Hence, smooth scheduling is far better than abrupt ones
in both efficiency and power.

\section{Discussion} 
\label{concl}

We have put forward a unified theory of HE based on a coherence-free periodically-modulated multilevel quantum-mechanical working fluid (WF).
The theory allows us to interpolate between 
two opposite limits of cycle scheduling (cycle forms): continuous and smoothed multi-stroke cycles (approximating any cycle, e.g. Otto, Carnot or two-stroke cycles). 

The following universal features emerge from this unified treatment: (a) The setup may operate as a heat engine, heat distributor, or a refrigerator, depending on the modulation 
rate $\Delta_{\rm m}$ and the smoothness parameter $k$. (b) The efficiency  increases for any cycle form with the WF frequency modulation-rate $\Delta_{\rm m}$, 
attaining a maximum which is bounded by the Carnot bound, before becoming ill-defined
 at the speed limit set by a 
modulation rate $\left(\Delta_{\rm m}\right)_{\rm SL}$, above which 
the setup stops acting as a heat engine. (c) Remarkably, the hitherto unexplored hybrid cycles may give rise to simultaneous dual action as refrigerator and engine. A 
conceptually novel modulation-induced power 
boost is predicted here for hybrid HE cycles. The reason is that hybrid cycles yield an optimal tradeoff between speed and the cost of coupling to the baths which is never turned 
off or on completely.

Despite these universal trends, the 
different scheduling (cycle) forms are not equivalent, but strongly depend on the ``smoothness'' parameter $k$ which accounts for the cost of coupling and decoupling from the baths. 
 The continuous cycle outperforms a  non-abrupt, realistic
Otto cycle (ROC) in terms of 
its maximal efficiency
near their respective speed limits, imposed by the condition  on the transition to the refrigerator regime. On the other hand, a hybrid cycle may outperform both the ROC 
and the continuous cycles in terms of the maximal 
power output.  Furthermore, the HE obeying a continuous or hybrid 
cycle  can operate  at a thermodynamic 
steady-state (TSS) that approximates a Gibbs state, while a finite-time ROC fails to reach a TSS (App. D). 
Such inequivalence of the cycle forms grants us the freedom to optimize the HE so that it attains maximal efficiency or power under fast modulation. 

Our analysis enables one to engineer a wide class of HEs, as long as $H_{\rm S}(t)$ and $H_{\rm I}(t)$ obey the commutative requirement Eq. (\ref{commi}) and are periodic with $\Delta_{\rm m}$, which is taken to be
an integral multiple of $\Delta_{\rm I}$. 
For example, one can have a global Hamiltonian in the form of Eqs. (\ref{hamilsb}) - (\ref{hamili}) for a Carnot cycle in
the limit $k\to\infty$ instead of a ROC, upon replacing  $\omega_{\rm Otto}$
by an
$\omega_{\rm Carnot}$ in Eq. (\ref{omteq}) in order to engineer isothermal expansion and compression strokes. On the other hand,
a realistic two-stroke cycle in the $k\to\infty$ limit would 
require two separate
WFs \ct{kosloff15} connected intermittently to only the cold or the hot bath with a similar parameterization of $f_{\rm j}(k,t)$ considered here, but with $f_{\rm c}(k,t) = f_{\rm h}(k,t)$ for all $k, t$, and an appropriate choice of 
$\omega(k,t)$ to include the interaction between the two WFs.

 The central issue of quantumness has thus been elucidated: the expressions for the efficiency and power bounds are independent of the WF quantized level 
structure and have classical counterparts, provided quantum coherence effects are absent. The foregoing features and the lack of 
quantum (coherence-assisted) 
effects are consequences of a driving Hamiltonian $H_{\rm S}(t)$ that commutes with itself at all times 
(see however bath-induced persistent quantum coherence in a degenerate multilevel WF \ct{brumer15}). By contrast, HE whose $H_{\rm S}(t)$ does not commute with itself 
\ct{kosloff84, geva92, feldmann00, rezek06, levy14} may face unwarranted friction effects.

These results and insights  suggest that traditional thermodynamics is adhered to, whereas quantum coherence is neither essential nor advantageous for HE performance. They map out all options for incoherently operating HE and may serve as guidelines for optimal 
HE designs based on quantum systems in various experimental scenarios \ct{abah14, alvarez10, rossnagel16}: (i) optomechanical HE machines \ct{ klimovsky15b} or (ii) HE based 
on multilevel WF e.g. a molecular rotator, modulated by electromagnetic 
fields and interacting with intra-cavity heat baths, (iii) HE based on a WF of Rydberg atoms. 

{\bf Acknowledgements:}  The authors acknowledge Raam Uzdin, Amikam Levy and Ronnie Kosloff for helpful comments and suggestions. The support of BSF, ISF, AERI and VATAT is acknowledged.

\section*{Appendices}
\setcounter{equation}{0}
\renewcommand{\theequation}{A\arabic{equation}}

\subsection{Floquet Analysis of the Master Equation}
\label{applv}
One can write the rate of change of the system density operator $\rho(t)$ in the interaction picture as
\ba
\dot{\rho}(t) &=& - \int^t_0 ds {\rm Tr}_{B}\big[f_c(k, t)\hat{S}(t)\otimes \hat{B}_c(t) \non\\ 
&+& f_h(k, t)\hat{S}(t)\otimes \hat{B}_h(t), [f_c(k,s)\hat{S}(s)\otimes \hat{B}_c(s) \non\\
&+& f_h(k,s)\hat{S}(s)\otimes \hat{B}_h(s), \rho(t)\otimes \rho_B]\big].
\label{rhosdot}
\ea 
In what follows, we focus on one of the baths and omit the labels $c/h$. We then have
\ba
\hat{S}^{\dagger}(t) &=& \hat{S}(t)\non\\
\hat{B}^{\dagger}(t) &=& \hat{B}(t)\non\\
{\rm Tr}\left[\hat{B}(t) \hat{B}(s) \rho_B\right] &=& \langle \hat{B}(t) \hat{B}(s) \rangle \equiv \Phi(t-s)\non\\
f(k, t) &=& \sum_{r = -N_{f}}^{N_f} f_r e^{-ir\Delta_{\rm I} t} \non\\
\hat{S}(t) &=& \sum_{q \geq 0,\omega} S_{\pm q\omega} e^{-i(\omega \pm  q\Delta_{\rm m})t},
\label{misc}
\ea
where $\Delta_{\rm I} = 2\pi/\tau_{\rm I}$, $\Delta_{\rm m} = 2\pi/\tau_{\rm m}$ and $\Delta_{\rm m} \geq \Delta_{\rm I}$. 

One can use Eq. (\ref{misc}) to write the first term on the r.h.s. of Eq. (\ref{rhosdot}) as
\ba
T_1 &=& - \sum_{r,r^{\prime},\omega, \omega^{\prime}, q \geq 0, q^{\prime}\geq 0} f_r f_{r^{\prime}} e^{-i(r+r^{\prime})\Delta_{\rm I} t} e^{-i[(\omega^{\prime} - \omega) \pm  (q^{\prime} - q)\Delta_{\rm m}] t} \non\\
&&\hat{S}^{\dagger}_{\pm q^{\prime}\omega^{\prime}}\hat{S}_{\pm q\omega} \rho(t)\int^{t}_0 \big[ \Phi(t-s)  e^{-i(\omega \pm  q\Delta_{\rm m})(t-s)} \non\\ 
&&e^{ir^{\prime} \Delta_{\rm I}(t-s)}\big] ds,
\label{rwaT1}
\ea
where we have assumed that $f(t)$ varies slowly so that $r$ is finite, and have taken into account $\Delta_{\rm I} \ll \Delta_{\rm m}$.

In the limit of large times,  the rotating wave approximation requires
\ba
\Delta_{\rm m} &=& \mp \frac{\omega^{\prime} - \omega}{q^{\prime} - q}, \non\\
\omega^{\prime} &=& \omega;~~q^{\prime} = q.
\label{rwaa}
\ea
Condition (\ref{rwaa}) 
gives us
\ba
T_1 &\approx& -\frac{1}{2} f(k,t)^2 \sum_{\omega, q\geq 0} \hat{S}^{\dagger}_{\pm q\omega}\hat{S}_{\pm q\omega} \rho(t) G(\omega,\pm q).
\label{rwaqq2}
\ea

We note that in the limit of Otto cycle ($k \to \infty$), $f_r$ can be non-zero for large $|r|$; further, $\Delta_{\rm m} = \Delta_{\rm I}$, implying Eq. (\ref{rwaqq2}) becomes invalid. 
However, one can still analytically solve the dynamics by writing separate 
master equations for the isentropic and isochoric strokes. Following similar consideration for the other terms, along with the Markovian approximation and the KMS condition, we get for
$S_{q \omega} = a(\omega = \omega_0)$ in the harmonic-oscillator WF
\ba
\mathcal{L}_{j,\pm q}(t)\rho &=& f_j(k, t)^2\frac{P_q}{2}[G_{j}(\omega_0 \pm  q\Delta_{\rm m})([a\rho, a^{\dagger}] + \left[a, \rho a^{\dagger} \right]) \non\\
&+& G_{j}(-\omega_0 \mp q\Delta_{\rm m})(\left[a^{\dagger}\rho, a\right] + \left[a^{\dagger}, \rho a \right])],
\label{lv}
\ea
where $j = c, h$ denotes the cold/hot baths, $\omega_0 = \frac{1}{\tau}\int^{\tau}_0 \omega(t) dt$, 
\ba
P_q &=& P_{-q} \non\\
&=&  \left|\frac{1}{\tau_{\rm m}} \int^{\tau_{\rm m}}_0 e^{i \int^{t}_0 \left(\omega(t) - \omega_0 \right) dt^{\prime}} e^{i q \Delta_{\rm m} t} dt\right|^2 \non\\
G_{j}(\omega_0 \pm  q\Delta_{\rm m}) &\equiv& G_{j}(\omega,\pm q) \non\\
&=& \int_{-\infty}^{\infty} e^{i(\omega_0\pm  q\Delta_{\rm m}) t} \left<\hat{B}_j(t) \hat{B}_j (0) \right> dt \non\\&=& e^{(\omega_0 \pm  q\Delta_{\rm m})/T_j} G_{j}(-\omega_0 \mp q\Delta_{\rm m}). \non
\ea
In order to allow HE operation, we require spectral separation of the baths by setting $G_{\rm h}(\omega_0 - q\Delta_{\rm m} < \omega_0) \approx 0$ and
$G_{\rm c}(\omega_0 + q\Delta_{\rm m} > \omega_0) \approx 0$.

\subsection{Cycle Parameterization}

We  parameterize $\omega(k,t)$ as shown in Eq. (\ref{omteq}). Here $\delta{\omega}_{\rm max}(k)$ gives the modulation amplitude of the input signal. 
For very small $\delta{\omega}_{\rm max}(k)$, the output power is also small, as is the case for
a continuous cycle $k = 0$. On the other hand, a ROC corresponds to large $\delta{\omega}_{max}(k)$, and hence large power. Therefore, in order to have a 
fair comparison of all the cycles, we have scaled the output power by $\delta{\omega}_{\rm max}(k)$ in Fig. (\ref{effk}).

\subsection{Heat Currents and Rate Equations}

The second law of thermodynamics gives us the generic expression for the heat currents \ct{klimovsky15}
\ba
J_j(t) = -\frac{1}{\beta_j}\sum_q {\rm Tr} \left[\mathcal{L}^j_q \rho(t) \ln \tilde{\rho}^{j q}  \right],
\ea
which, when combined with the rate equations (\ref{popls}), yields Eq. (\ref{Jhc}). Here $\mathcal{L}^j_q$ denotes the Lindblad operator corresponding to the $j$th bath and the $q$th mode, and 
$\tilde{\rho}^{j q}$ is the corresponding stationary state.

Equation (\ref{Jhc}) is expressed in terms of
\ba
\mathcal{F}_{\rm h}(q\geq0, k,t) &=& \sum_n n\Big(R_{\rm ne}(k,t) \\
&+& e^{-\beta_{\rm h}(\omega_{0} + q\Delta_{\rm m})}R_{\rm na}(k,t) \Big), \non\\
R_{\rm ne}(k,t) &=& G_0\left[(n+1)\mathcal{P}_{n+1}(k, t) - n\mathcal{P}_n(k, t)\right], \non\\
R_{\rm na}(k,t) &=& G_0\left[n\mathcal{P}_{n-1}(k, t) - (n+1) \mathcal{P}_n(k, t)\right]\non,
\ea
where  $n$ labels the WF levels and $R_{\rm ne}$ and $R_{\rm na}$ denote the $n+1 \to n$ 
(emission) and $n-1 \to n$ (absorption) rates, respectively.  For simplicity, they are assumed to scale with $G_0$, where
\ba
G_{\rm c}(0< \omega_0 - q\Delta_{\rm m} < \omega_0) &=& G_0;\non\\
G_{\rm h}(\omega_0 + q\Delta_{\rm m} > \omega_0) &=& G_0.
\ea

Let us sketch the derivation of these expressions. Inter-level coherences decay exponentially with time, such that at large 
times the evolution is given by the Pauli master equation for the WF level populations $\mathcal{P}_n(t) = \bra{n} \rho(t) \ket{n}$ \ct{breuer02}.
For a generic multi-level system with 
average energy spacing between the $n$th and $(n-1)$th levels given by $\tilde{\omega}_n$, the above derivation yields the population dynamics as
\begin{multline}
\frac{d}{dt} \mathcal{P}_{n}(k,t) =  G_0 \sum_{j,q\geq0} P_q(k) f_j^2(k, t)[(n+1) \non\\
\big(\mathcal{P}_{n+1}(k,t) - e^{-\beta_j (\tilde{\omega}_{n+1} \pm q\Delta_{\rm m})}\mathcal{P}_n(k,t)\big) \\
+ n\big(-\mathcal{P}_n(k,t) + e^{-\beta_j (\tilde{\omega}_n \pm q\Delta_{\rm m})} \mathcal{P}_{n-1}(k,t) \big)],
\end{multline}
where the plus (minus) sign corresponds to the hot (cold) bath. For a harmonic oscillator with equidistant energy levels $\tilde{\omega}_{n} = \omega_0$ for all $n$, this equation reads
\begin{multline}
\frac{d}{dt} \mathcal{P}_n(k,t) = G_0 \sum_{j,q\geq0} P_q(k) f_j^2(k, t)\\
\Big\{\big[(n+1)\mathcal{P}_{n+1}(k,t) - n\mathcal{P}_n(k,t)\big]  \\ 
+ e^{-\beta_j (\omega_0 \pm q\Delta_{\rm m})}\big[n\mathcal{P}_{n-1}(k,t) - (n+1)\mathcal{P}_n(k,t) \big]\Big\},\\
= \sum_{j,q\geq0} P_q(k) f_j^2(k, t)\left[R_{ne}(k,t) + e^{-\beta_j \omega_j} R_{na}(k,t)\right] 
\label{popls}
\end{multline}
where we have considered the KMS condition and taken into account that only the frequencies $\omega_j = \omega_0 - q\Delta_{\rm m}$ ($\omega_j = \omega_0 + q\Delta_{\rm m}$) contribute for the cold (hot) bath.

\subsection{Thermodynamic steady-state of the HE} 
\label{tess}

\begin{figure}[t]
\begin{center}
\includegraphics[width= \columnwidth, angle = 0]{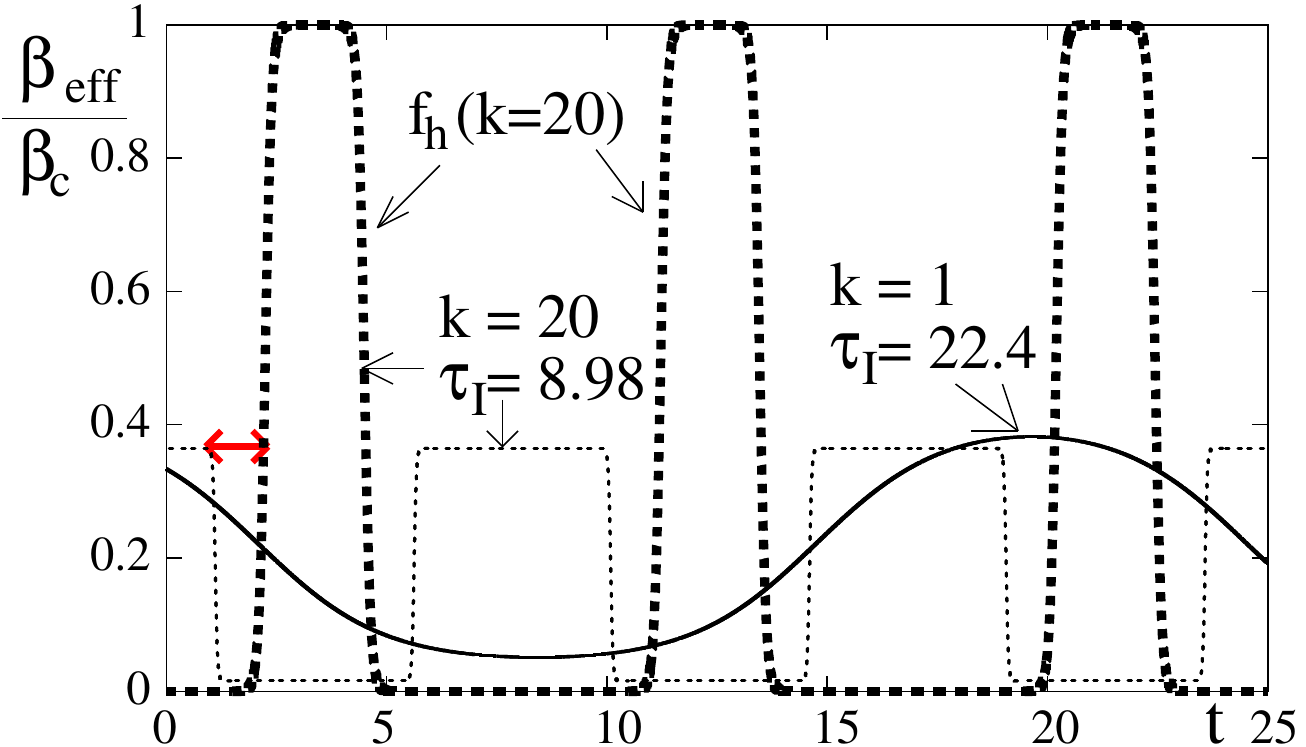}
\end{center}
\caption {(Color Online) $\beta_{\rm eff}/\beta_{\rm c}$ plotted as a function of time for $k = 1$ (solid line) and $k = 20$ (thin dotted line).
$\beta_{\rm eff}$ changes abruptly during the unitary stroke for $k = 20$, even though $f_h(t)$ (thick dashed line) $\approx 0$ for $t < \tau_{\rm I}/4$ (see red line for the discrepancy),
showing that $\beta_{\rm eff}$ loses its meaning in the limit of large $k$. By contrast, $\beta_{\rm eff}$ varies smoothly 
 for $k = 1$ (solid line).
(Same parameters as in figs. (\ref{eff_Del}, \ref{effk}), and $\Delta_{\rm m} = 1.4$).}
\label{betef}
\end{figure}
While $\left(\Delta_{\rm m}\right)_{\rm SL}$ marks the transition from a heat engine to a refrigerator, there can be another speed limit
that corresponds to onset of the thermodynamic steady state 
(TSS) as a limit cycle. A TSS must fulfill the condition of a slowly varying steady state $\rho^{\rm ss}(t)$, i.e. at any 
chosen initial time  $t_0$, it must satisfy
\ba
\dot{\rho}(t_0) = 0 \simeq \dot{\rho}^{\rm ss}(t_0) = \sum_n \dot{\mathcal{P}}^{\rm ss}_n(t_0) \ket{n}\bra{n},
\label{sstate}
\ea
under the initial condition $\rho(t_0) = \rho^{\rm ss}(t_0)$, $\mathcal{P}_n(k,t) = \langle n | \rho(k,t) | n\rangle$ being the WF level populations. 

The probability of occupation 
\ba
\mathcal{P}_n^{\rm ss}(t) = \frac{1}{\mathcal{N}}\left[\frac{f_{\rm h}^2(k, t) A_h + f_{\rm c}^2(k, t) A_c}{\left(f_{\rm h}^2(k, t) + f_{\rm c}^2(k, t)\right)\sum_{q>0} P_q}\right]^n
\ea
of the $n$th energy level in its instantaneous steady state follows the equation
\ba
\dot{\mathcal{P}}_n^{\rm ss}(t_0 + \Delta t) \approx \dot{\mathcal{P}}_n^{ss}(t_0) + \Delta t \ddot{\mathcal{P}}_n^{ss}(t_0)
\ea
for any arbitrary time $t_0$.
On the other hand, the instantaneous probability of occupation $\mathcal{P}_n(t)$ of the $n$th energy level evolves following Eqs. (\ref{lv1}) and (\ref{mastersq}) as
\ba
\dot{\mathcal{P}}_n(t_0 + \Delta t) \approx \dot{\mathcal{P}}_n(t_0)|_{ss} + \Delta t \ddot{\mathcal{P}}_n(t_0) =  \Delta t \ddot{\mathcal{P}}_n(t_0), \non
\ea
where we have assumed that the system is in its instantaneous steady state at time $t_0$, with $\dot{\mathcal{P}}_n(t_0)|_{ss} = 0$. Clearly, for the system to remain close 
to its instantaneous steady state at all times, 
\ba
\dot{\mathcal{P}}_n^{\rm ss}(t) \ll \Delta t \ddot{\mathcal{P}}_n^{\rm ss}(t_0)~\forall~t,
\ea
i.e., $\dot{\mathcal{P}}_n^{\rm ss}(t)$ need to be small. 
The thermodynamic steady state and hence $\beta_{\rm eff}$ become ill-defined in the limit of large $k$ (see Fig. \ref{betef}).

From Eq. (\ref{sstate}) we can derive the following estimate for the $k$-scaling of the time-scale $\tau_{\rm TSS}$ that describes the
steady-state variation of any $n$-state population $\mathcal{P}^{\rm ss}_n(t)$:
\ba
\tau_{\rm TSS} &\simeq& 4\pi \left|\left[\frac{f_{\rm c}(k, t) f_{\rm h}(k, t)\left(f_{\rm c}(k, t) - f_{\rm h}(k, t)\right)}{(f_{\rm c}^2(k,t) + f_{\rm h}^2(k,t))^2}\right] \zeta\right|_{\rm max} \non\\
\zeta &=& \frac{\left|e^{-\beta_{\rm h} \omega_2} - e^{-\beta_{\rm c} \omega_1}\right|}{\Delta t |\ddot{\mathcal{P}}_n^{\rm ss}(t)|},
\label{taumin}
\ea
for any choice of small $\Delta t \ll \tau_{\rm I}$ characterizing the deviation of $\dot{\rho}(t)$ from $\dot{\rho}^{\rm ss}(t)$  (App. C).  
The factor in the brackets denotes the maximum value determined only by the $f_j(k,t)$ factors.  In the continuous-cycle limit of 
$f_{\rm c}(k, t) = f_{\rm h}(k, t) = 1 ~ \forall t$, $\dot{\mathcal{P}}_n^{\rm ss}$ in (\ref{sstate}) is time independent so that TSS is achieved for any $\tau_{\rm I}$. On the other hand,
a TSS cannot be achieved in the 
 ROC limit for any finite cycle ($\tau_{\rm I} < \infty$) since the numerator and the denominator in the $k$-dependent part of (\ref{taumin}) vanish in the {\it isentropic} part of the cycle where
 $f_{\rm h} = f_{\rm c} = 0$. 
 
If we require the TSS to approximate a thermal 
(Gibbs) state with effective (positive) inverse temperature $\beta_{\rm eff}(k,t)$, it must obey the following 
relation for the ratio 
of adjacent level populations (cf. Eqs. (\ref{sstate}), (\ref{taumin}))
\ba
\label{beff}
&&\beta_{\rm eff}(k,t)  \equiv -\frac{\log\left[\frac{f_{\rm h}^2(k, t) A_{\rm h} + f_{\rm c}^2(k, t) A_{\rm c}}{\left(f_{\rm h}^2(k, t) + f_{\rm c}^2(k, t)\right)\sum_{q>0} P_q}\right]}{\omega_0}, \\
&& A_{\rm h} = \sum_{q>0}P_{q}e^{-\beta_{\rm h} (\omega_0 + q\Delta_{\rm m})};~ A_{\rm c} = \sum_{q>0}P_{q}e^{-\beta_{\rm c} (\omega_0 - q\Delta_{\rm m})}\non.
\ea

Eq. (\ref{beff}) cannot hold for $k \to \infty$, since  the $k$-dependent factor in Eq. (\ref{taumin})  becomes ill-defined when $f_{\rm c} = f_{\rm h} = 0$ in 
the {\it unitary stroke}. This 
results in unphysical 
behavior of $\beta_{\rm eff}$; one can show that $\beta_{\rm eff}$ abruptly changes during the unitary (isentropic) stroke, where it should be constant, owing to the baths being decoupled from the WF during this stroke. 
Hence, a finite-time  ROC
 is not amenable to a Gibbs-state description, as opposed to hybrid and continuous cycles. 
\end{document}